\documentclass[10pt]{IEEEtran}
\usepackage{stfloats}

\usepackage{enumerate}
\usepackage{algorithm,algpseudocode}
\usepackage{setspace,amsmath,latexsym,cite,amssymb,epsfig,amsfonts}
\usepackage{url,cite}
\usepackage{graphicx}
\usepackage{psfrag}
\usepackage{footmisc}
\usepackage{multirow}
\usepackage{color}
\usepackage{multicol}
\usepackage{mathtools}
\usepackage{subcaption}
\usepackage{graphicx}
\usepackage{amssymb}
\usepackage{amsmath}
\usepackage{epstopdf}
\usepackage{geometry}
\usepackage{float}
\usepackage{balance}
\usepackage{amsthm}

\algnewcommand\algorithmicforeach{\textbf{for}}
\algdef{S}[FOR]{For}[1]{\algorithmicforeach\ #1\ \algorithmicdo}

\twocolumn

\theoremstyle{remark}

\newcommand{\bfg}[1]{\mbox{\boldmath$#1$}}
\newcommand{\tra}{{\mathrm{\scriptscriptstyle T}}}

\newcommand{\s}{{\mathbf s}}

\makeatother

        \makeatletter
        \def\fps@eqnfloat{!t}
        \def\ftype@eqnfloat{4}
        
        \newenvironment{eqnfloat*}
               {\@dblfloat{eqnfloat}}
               {\end@dblfloat}
        \makeatother

\title{Link Priority Buffer-Aided Relay Selection with Energy Storage from Energy Harvest}
\author{Mohammad Alkhawatrah, Yu Gong, Chong Huang and Gaojie Chen
\thanks{Mohammad Alkhawatrah is with Communications and Computer Engineering Department, Al-Ahliyya Amman University, Amman, Jordan, Email: {m.alkhawatrah@ammanu.edu.jo}}
\thanks{Yu Gong is with Wolfson School of Mechanical, Electrical and Manufacturing Engineering, Loughborough University, UK, Email: { y.gong@lboro.ac.uk}. Corresponding author: Yu Gong.}
\thanks{Chong Huang and Gaojie Chen are with 5GIC \& 6GIC, Institute for Communication Systems (ICS), University of Surrey, United Kingdom Email: {\{chong.huang, gaojie.chen\}@surrey.ac.uk.}}
}

\begin{document}
\captionsetup[figure]{name={Fig.},labelsep=period}

\begin{singlespace}
\maketitle
\end{singlespace}

\thispagestyle{empty}
\begin{abstract}
This paper proposes a novel relay selection scheme for buffer-aided wireless networks with relays equipped with both data buffers and energy storage. While buffer-aided relay networks have demonstrated significantly improved performance, energy harvesting has become an attractive solution in many wireless systems, garnering considerable attention when applied to buffer-aided relay networks. It is known that state-dependent selection rules must be used to achieve full diversity order in buffer-aided relay networks, requiring link priorities for data transmission to be set based on system states. This task becomes challenging when both data buffers and energy storage are involved. In this paper, we introduce a novel method for setting link priorities, which forms the basis for a new selection rule. The outage probability of the proposed selection scheme is derived. The simulation results demonstrate the superiority of our proposed algorithm which achieves full diversity in buffer-aided relay selection with energy storage, and consistently outperforms baseline approaches across various metrics.

\end{abstract}


\section{Introduction}
\IEEEPARstart{A}{s} wireless communication rapidly advances, the buffer-aided cooperative network is receiving significant attention in future beyond fifth generation (B5G) and sixth generation (6G) research due to its considerable potential \cite{9363027,10500738}. This paper investigates relay selection in wireless networks where relays are equipped with both data buffers and energy storage. A typical relay network (which can be found in various scenarios such as the 4G/5G/6G mobile networks, Internet-of-Things and sensor networks) consists of one source node, one destination node, and $K$ relay nodes. By selecting one relay for data transmission at a time, the network can effectively harvest diversity gain \cite{Su2010}. Initially, the max-min relay selection was considered optimal, but later research found that its performance is limited by the concurrent selection of source-to-relay and relay-to-destination links \cite{Bletsas20063}. This limitation can be overcome by incorporating data buffers at the relays.

Two typical early approaches for buffer-aided relay networks are the max-max and max-link schemes. In the max-max scheme, the source-to-relay and relay-to-destination links are alternately selected for data transmission, but not necessarily via the same relay node \cite{Ikhlef201203}. This scheme can achieve a 3dB coding gain but maintains the same diversity order of $K$ as the max-min scheme. The max-link scheme selects from all available source-to-relay and relay-to-destination links, reaching a full diversity order of $2K$ when the buffer size is large \cite{Krikidis20125}. However, the diversity order decreases rapidly with smaller buffer sizes. Both the max-max and max-link schemes, including their variants, do not consider the states of the buffers (i.e., the number of packets in the buffers). This oversight can easily result in the buffers becoming either empty or full, thereby rendering the corresponding source-to-relay or relay-to-destination links unavailable for data transmission. Consequently, this leads to a reduced diversity order.

To further explore performance improvements in cooperative networks, state-dependent schemes have been introduced to select links based not only on channel coefficients but also on buffer states. This is accomplished by assigning higher priority, either explicitly or implicitly, to links that are less likely to cause buffers to become empty or full. For instance, in \cite{Luo2015}, a state-based relay selection method is proposed to maintain the buffer length at two, thereby minimizing the risk of buffer underflow or overflow. This approach effectively prioritizes links that are more likely to keep the buffer length at two. In \cite{GongSocial2018}, link priorities are explicitly determined based on buffer states, enabling a trade-off between delay and throughput performance. Additionally, to leverage the impact of outage probability in buffer-aided relay selection, a probabilistic scheme was proposed to achieve a balance between outage probability and average delay in cooperative networks \cite{9006913}. In \cite{9933941}, a novel buffer-aided relay selection strategy was studied to adaptively coordinate the selection priorities for direct links and relay links between the source and the destination. In \cite{9321455}, a decision-assisted reinforcement learning-based selection method was proposed to balance delay and throughput performance in buffer-aided cooperative networks, where the priorities of the links are obtained automatically during the learning process. In \cite{9337192}, acknowledgment packets were utilized to report the shared information in the state-based relay selection scheme.

Buffer-aided relay selection has been investigated in various applications, including the space-time coding \cite{Peng2016}, multi-hop relay transmission \cite{GongHop2016}, non-orthogonal-multiple-access (NOMA) system \cite{Zhang2017}, the Internet-of-Things (IoT) \cite{9281119}, cognitive networks \cite{9367290}, secure communications \cite{9536592}, satellite communications \cite{10436093} and reconfigurable intelligent surface \cite{9508416}. These approaches typically assume that relay nodes have access to a fixed power supply. However, in scenarios where a power supply is difficult to obtain, such as many IoT applications \cite{Sun2018}, wireless energy harvesting provides an attractive solution. Applying energy harvesting in buffer-aided relay selection has garnered significant attention. For instance, in \cite{Ahmed2014}, optimal power allocation is proposed for a three-node buffer-aided link selection with energy harvesting. In \cite{LiuY2018}, adaptive buffer-aided relay communication with energy storage is explored in a three-node relay network. Additionally, a buffer-aided multiple relay selection with unit-sized energy storage is proposed in \cite{Krikidis2015}. Early buffer-aided relay selection schemes with energy harvesting typically considered either a single relay node or unit-sized energy storage. In \cite{LiuK2021}, a multiple relay network with finite sizes for both data buffers and energy storage is examined, assuming relay nodes only harvest wireless energy from source node transmissions. A more general scenario is considered in \cite{HuangS2022}, where relay nodes harvest energy from both source and relay transmissions. In a more recent work \cite{10130605}, reinforcement learning is employed to minimize the average age of information (AoI) in a buffer-aided relay network with energy harvesting, where only a unit buffer size at the relays is considered.

The selection rules in both \cite{LiuK2021} and \cite{HuangS2022} follow a principle similar to that in the max-max scheme, where source-to-relay and relay-to-destination links are selected alternately. Consequently, all issues inherent in the max-max scheme remain. First, the diversity order can only reach up to $K$, which is half of the full diversity order. Secondly, the selection process can easily encounter deadlock, meaning no link can be selected regardless of channel strength. In the max-max scheme, deadlock occurs when the transmission is in the source-to-relay cycle but all buffers are full, or when it is in the relay-to-destination cycle but all buffers are empty. To handle deadlock, the max-max scheme suggests pre-filling the buffers to half capacity to reduce such events, which limits the selection's flexibility. In \cite{LiuK2021} and \cite{HuangS2022}, the deadlock issue becomes more severe due to the addition of energy storage. The proposed schemes switch to max-min when deadlocks occur, but this results in a loss of the benefits provided by the buffers.

To achieve full diversity order, state-dependent relay selections must be applied, which depend on how link priorities are set. Without energy harvesting, the selected link only affects the buffer length of the corresponding relay, while the buffers of other relays remain unchanged. Thus, the priority of the link can be easily set based on the current buffer state. However, when energy storage is applied, the selected link affects not only the data buffer of the corresponding relay but also the energy storage of other nodes. Since an empty energy storage also renders the corresponding relay-to-destination link unavailable for data transmission, it becomes challenging to set link priorities based on the current state, which now includes both data buffers and energy storage.

This challenge is likely the main reason that in \cite{LiuK2021} and \cite{HuangS2022}, the schemes follow the max-max approach, where source-to-relay and relay-to-destination links are selected alternately. While this simplifies setting link priorities, it sacrifices the full diversity order.

As far as the authors are aware, no existing buffer-aided relay selection with energy harvesting has achieved a full diversity order of $2K$. The main contributions of this paper are summarized as follows:

\begin{itemize}
\item Novel link priority approach: We propose a novel method for setting link priorities during selection. Similar to the scheme in \cite{GongSocial2018}, the link priorities are explicitly determined based on both the channel gains and the system states. However, unlike \cite{GongSocial2018}, our approach incorporates both the data-buffer status and the energy storage status into the system state.

\item New link selection strategy: Based on the proposed link priority mechanism, a new link selection strategy is introduced. Similar to the max-link scheme, the link is selected from among all available source-to-relay and relay-to-destination links. Particularly the proposed strategy introduces the concept of "edge-state" to prevent the system from entering a deadlock, ensuring smoother operation and improved reliability.

\item Closed-form outage probability: We derive a closed-form expression for the outage probability and analyze the diversity order. Our analysis demonstrates that the proposed scheme can achieve a full diversity order of $2K$ when a sufficient number of relays are available. To the best of the authors' knowledge, this is the first scheme capable of achieving full diversity in buffer-aided relay selection with energy storage considerations.

\item Numerical simulations: We conduct numerical simulations to validate the effectiveness of the proposed scheme. The results show that our approach outperforms baseline methods across various performance metrics, highlighting its practical advantages.
\end{itemize}

The rest of the paper is organized as follows: Section \ref{sec:model} describes the system model; Section \ref{sec:ava} proposes a novel approach for assigning priorities for link selection; Section \ref{sec:rule} describes the selection rules based on the priorities obtain in the previous section; Section \ref{sec:out} derives the closed-form for the outage probability and analyze the diversity order; Section \ref{sec:num} presents numerical results to verify the proposed scheme.; Finally Section \ref{sec:con} concludes the paper.


\section{System model}
\label{sec:model}

\begin{figure}[h]
  \centering
  \centerline{\includegraphics[scale=0.5]{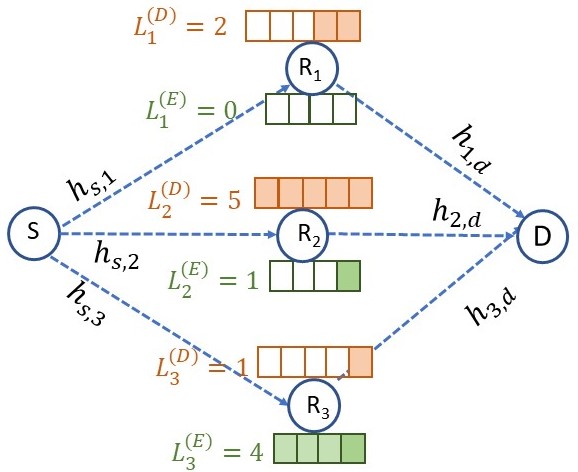}}
 \caption{\small The 3-node relay network with fixed size data buffer and energy storage.} \label{fig:example}
\end{figure}

We consider a relay network with one source node $S$, one destination node $D$ and $K$ relays $R_k$ ($k \in  \{1, 2, ..., K\}$). Every relay, which applies the decode-and-forward, is equipped with one data buffer of size $L^{(D)}_{{\max}}$ and one energy storage of size $L^{(E)}_{{\max}}$. The 3-relay case is illustrated in Fig. \ref{fig:example}. The channel coefficients for $S\to R_k$ and $R_k \to D$ links are denoted as $h_{s\to k}$ and $h_{k\to d}$, respectively. We assume the channels are block fading that they keep constant within one time slot, but vary from one time slot to another. Moreover, we assume the link between $S$ and $D$ is unavailable due to severe blocking or deep fading \cite{9448155}. For later use, we denote $link_{s \to k}$ and $link_{k\to d}$ to represent the $S \to R_k$ and $R_k \to D$ links, respectively. Particularly, $link_k$ refers to either $link_{s \to k}$ or $link_{k\to d}$.

The transmit powers at source and relay nodes are $P_s$ and $P_r$, respectively. We let $P_r = \alpha P_s$, where $0 < \alpha  \leq 1$ which is the relay power coefficient.

Suppose at time slot $t$, the length of the data buffer and energy storage of the relay $R_k$ are $L^{(D)}_{k,t}$ and $L^{(E)}_{k,t}$, respectively. If $link_{s\to k}$ is selected for data transmission, the data buffer length of $R_k$ is increased by one that
\begin{equation}
    L^{(D)}_{k,t+1} = \min\{ L^{(D)}_{k,t} + 1, ~~L^{(D)}_{\max} \}
\end{equation}
At the same time, the other relays harvest energy from the wireless transmission at $S$. The energy harvested at relay $R_j$ during one time slot $T$ is given by
\begin{equation}
    \Delta E_{j,t} = \rho \cdot P_s \cdot |h_{s\to j,t}|^2 \cdot T, \quad j\neq k
\label{eq:DeltaE}
\end{equation}
where $\rho$ is the energy harvesting coefficient.

For every packet transmission at the relays, it consumes $E_r = P_r  T = \alpha P_s T$ energy. Every unit in the energy storage contains $E_r$ energy. We let $m = \lfloor \Delta E_{j,t}/E_r \rfloor = \lfloor (\rho/\alpha) |h_{s\to j,t}|^2  \rfloor$, where $\lfloor. \rfloor$ rounds the value to the lowest integer. Thus if $link_{s\to k}$ is selected, the length of the energy storage at $R_j$ is increased by $m$ that
\begin{equation}
    L^{(E)}_{j,t+1} = \min\{ L^{(E)}_{j,t} + m, ~~L^{(E)}_{\max} \}
\end{equation}

On the other hand, if $link_{k\to d}$ is selected for data transmission at time slot $t$, the data buffer length at $R_k$ is decreased by one that
\begin{equation}
    L^{(D)}_{k,t+1} = \max\{ L^{(D)}_{k,t} - 1, ~~0 \}
\end{equation}
The length of the energy storage at $R_k$ is also decreased by one that
\begin{equation}
    L^{(E)}_{k,t+1} = \max\{ L^{(E)}_{k,t} - 1, ~~0 \}
\end{equation}
The data buffer and energy storage for other relays remain unchanged.


The lengths of the data buffer and energy storage at all relays form the system state. To be specific, at time $t$, the state vector $\s(i)$ is given by
\begin{equation}
    \s(i) = [\s^{(D)}(i), ~\s^{(E)}(i)],
\label{eq:std}
\end{equation}
where $ \s^{(D)}(i)$ is the data state vector as
\begin{equation}
 \s^{(D)}(i) = [L^{(D)}_1(i), \cdots, L^{(D)}_K(i)]
\end{equation}
and $\s^{(E)}(i)$ is the energy state vector as
\begin{equation}
    \s^{(E)}(i) = [L^{(E)}_1(i), \cdots, L^{(E)}_K(i)]
\end{equation}
The time index $t$ will be ignored in below selections without causing confusion.

We note that in this paper, we assume that relay nodes only harvest energy from source node transmissions. This assumption is made for the sake of clarity. The relay selection method proposed in the following sections can be easily extended to the scenario where relay nodes harvest energy from all transmissions.


\section{Priorities for link selection}
\label{sec:ava}

The number of available links for selection is crucial for relay selection. The more available links there are, the less likely it is for the transmission to be in outage. In a network with $K$ relays, the maximum number of available links is $2K$, consisting of $K$ source-to-relay links and $K$ relay-to-destination links. A source-to-relay link becomes unavailable when the data buffer is full, and a relay-to-destination link becomes unavailable when either the data buffer or energy storage is empty. To achieve better outage performance, links that potentially lead to more available links should be given higher priority in any state.

\subsection{State availability-vector}
\label{sec:stavec}

The availability-vector is introduced to measure how available the links are at state $\s(i)$. Suppose at state $\s(i)$, the lengths of the data buffer and energy storage at relay $R_k$ are $L^{(D)}_k(i)$ and $L^{(E)}_k(i)$, respectively. The {\em availability-index for empty data buffer} for $R_k$ at $\s(i)$ is defined as how far the data buffer is away from empty which is given by
\begin{equation}
    M^{(D-empty)}_k(i) = L^{(D)}_{k}(i)
\label{eq:MDe}
\end{equation}
The {\em availability-index for full data buffer} is defined as how far the data buffer is away from full which is given by
\begin{equation}
    M^{(D-full)}_k(i) = L^{(D)}_{\max} - L^{(D)}_{k}(i)
\label{eq:Df}
\end{equation}
The {\em availability-index for empty energy storage} is defined as how far the energy storage is away from empty which is given by
\begin{equation}
    M^{(E-empty)}_k(i) = L^{(E)}_{k}(i)
\label{eq:Ed}
\end{equation}
Stacking \eqref{eq:MDe} \eqref{eq:Df} and \eqref{eq:Ed} together for all relays gives the availability-vector at state $\s(i)$ as
\begin{equation}
\begin{aligned}
    \mathbb{M}^{(\s(i))} &= \left [ M^{(D-empty)}_k(i), M^{(D-full)}_k(i), M^{(E-empty)}_k(i) \mid \forall k  \right ]  \\
    &= \left [  L^{(D)}_k(i), ~L^{(D)}_{\max} - L^{(D)}_{k}(i), ~L^{(E)}_{k}(i) \mid \forall k  \right ]
\end{aligned}
\label{eq:mMsi}
\end{equation}

For the example in Fig. \ref{fig:example}, the data buffer size $L^{(D)}_{\max}=5$, the maximum energy length $L^{(E)}_{\max}=4$, and the state vector is given by
\begin{equation}
\begin{aligned}
    \s &= [(L^{(D)}_{1}, \cdots, (L^{(D)}_{K}), (L^{(E)}_{1}, \cdots, L^{(E)}_{K})] \\
    &= [(2, 5, 1), (0, 1, 4)]
\end{aligned}
\end{equation}
The availability indices for every relay at the above state is shown in Table I.

\begin{table}[h]
\begin{center}
\begin{tabular}{ |c|c|c|c| }
 \hline
  & $ M^{(D-full)}_k$ & $ M^{(D-empty)}_k$ & $ M^{(E-empty)}_k$  \\
 \hline
 $R_1$ & 3 & 2 & 0 \\
 \hline
 $R_2$ & 0 & 5 & 1 \\
 \hline
 $R_3$ & 4 & 1 & 4 \\
 \hline
\end{tabular}
\end{center}
\caption{Availability indices for the example in Fig. \ref{fig:example} }
\end{table}
The availability-vector for the state shown in Fig. \ref{fig:example} is then obtained by stacking all indices in Table I, from the smallest to the largest for better exposition, into a vector as
\begin{equation}
 \mathbb{M}^{(\s)} = \left [0, 0, 1, 1, 2, 3, 4, 4, 5  \right ]
\label{eq:Mst0}
\end{equation}

The link availability at state $\s(i)$ is determined by the minimum index in $\mathbb{M}^{(s(i)))}$. The smaller the value, the less availability there is for the links in that state. If two states have the same minimum index in the availability-vector, the second minimum index is then compared and so on. If $ \mathbb{M}^{(\s(i))}$ shows higher availability than  $\mathbb{M}^{(\s(j))}$, it is denoted as
\begin{equation}
    \mathbb{M}^{(\s(i))} \succ \mathbb{M}^{(\s(j))}
\end{equation}


\subsection{Link priorities}
\label{se:linkpri}

The availability-vector is used to give link priorities for selection. Suppose at state $\s(i)$, there are $N_i$ available links. The priorities for the $N_i$ available links are set as follows:
\begin{itemize}
    \item For every available $link_k$, if it is selected, find the state $\s(k)$ to which $\s(i)$ will transit.

    \item From \eqref{eq:mMsi}, obtain the availability-vector $\mathbb{M}^{(\s(k)})$ of $\s(k)$.

    \item For any pair of available links, $link_{k_1}$ and $link_{k_2}$, respectively. If $\mathbb{M}^{(\s(k_1))} \succ \mathbb{M}^{(\s(k_2))}$, $link_{k_1}$ is given higher priority than $link_{k_2}$ for link selection.
 \item If $\mathbb{M}^{(\s(k_1))} = \mathbb{M}^{(\s(k_2))}$, the priorities of $link_{k_1}$ and $link_{k_2}$ are randomly set between them.
\end{itemize}


\subsection{An example}

For illustration, we look at the example in Fig. \ref{fig:example}, where we assume the energy harvest ratio $\rho=0.5$, the transmit power at $S$ is $P_s=1$, the relay power coefficient $\alpha=1$, the average gains for all channels are the same which are $E(|h_{k}|^2)=2$, and the time slot $T=1$. Suppose at time $t$, as is shown in Fig. \ref{fig:example}, the state is given by
\begin{equation}
   \s(i) = [(2, 5, 1), (0, 1, 4)]
\end{equation}

At state $\s(i)$, $link_{1 \to d}$ and $link_{s\to 2}$ are unavailable for link selection due to empty energy storage and full data buffer, respectively. The other links are available which will be considered one-by-one.

If $link_{s \to 1}$ is selected, the data buffer length at $R_1$ is increased by one, and the energy storage at $R_2$ and $R_3$ are expected to increase by $\Delta E_{k}(j)= \rho \cdot P_t \cdot {\rm E}(|h_{s \to k}|^2) = 1$ such that the corresponding energy storage will be increased by one. Then the state will transit to
\begin{equation}
    \s(j) = [(3, 5, 1), (0, 2, 4)]
\end{equation}

Following the similar procedure as that in obtaining \eqref{eq:Mst0}, the availability-vector at $\s(j) $ can be obtained as
\begin{equation}
 \mathbb{M}^{(s(j))} = \left \{ 0, 0, 1, 2, 2, 3, 4, 4, 5  \right \}
\end{equation}

Applying this process on every available link, we can obtain the availability-vectors for each of the corresponding state being transited, which are shown in Table \ref{tab:stallh}.

\begin{table}[h]
\begin{center}
\begin{tabular}{ |c|c|}
 \hline
  & $ \mathbb{M}^{(s(j))}$  \\
 \hline
 $h_{s,1}$ & \{ 0, 0, 1, 2, 2, 3, 4, 4, 5  \} \\
 \hline
 $h_{s,3}$ & \{ 0, 1, 2, 2, 2, 3, 3, 4, 5  \} \\
 \hline
 $h_{2,d}$ & \{ 0, 0, 1, 1, 2, 3, 4, 4, 4  \} \\
 \hline
 $h_{3,d}$ & \{ 0, 0, 0, 1, 2, 3, 3, 5, 5  \} \\
 \hline
\end{tabular}
\end{center}
\caption{Availability of the new state due to the selection of the corresponding links.}
\label{tab:stallh}
\end{table}

\begin{figure}[h]
  \centering
  \centerline{\includegraphics[scale=0.5]{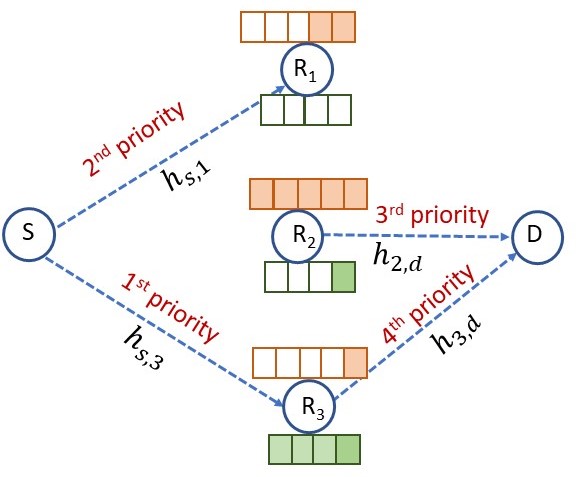}}
 \caption{\small The priorities of the available links in the example in Fig. \ref{fig:example}.} \label{fig:selection}
\end{figure}

From Table \ref{tab:stallh}, the priorities for all available links at $\s(i)$  are obtained as, from the highest to the lowest priority, $link_{s \to 3}$, $link_{s \to 1}$, $link_{2 \to d}$ and $link_{3 \to d}$, , respectively. This is shown in Fig. \ref{fig:selection}.


\section{Selection rules}
\label{sec:rule}

The selection rules are designed to prevent states with empty or full data buffers, as well as empty energy storage, as much as possible. As shown in the previous section, this is achieved by maintaining the system in states with higher link availability which is measured by $\mathbb{M}^{(\s)}$.

Below we will first introduce the `edge-state', and describe the selection rules for the edge-states and non edge-states, respectively.


\subsection{Edge state}

Unlike the max-link scheme (or its state-dependent variants) without energy harvesting, the simultaneous use of data buffers and energy storage can lead to a deadlock. This occurs when all data buffers are full and all energy storage units are empty. In such a scenario, no links can be selected, regardless of channel strength, resulting in a deadlock.

To address this, we introduce the concept of the `edge state'. In an edge state, all energy storage units are empty, one data buffer is nearly full, and all other data buffers are completely full. For the network with $K$ relays, there are $K$ edge-states out of a total of $[(L^{D}_{\max}+1)\times (L^{E}_{\max}+1) ]^K$ possible states.

Suppose that at edge state $\s(i)$, only the data buffer of relay $R_k$ is nearly full, while all other data buffers are completely full. In this situation, if the only available link, $link_{s \to k}$, is selected but the energy harvested is insufficient to charge any other relay, all data buffers will become full, and all energy storage units will remain empty. Consequently, the system will enter a deadlock.

To prevent the system from entering a deadlock, the selection rule is to select $link_{s \to k}$ only if the following two conditions are both met:
\begin{itemize}
    \item The capacity of $link_{s \to k}$ is larger than the target data rate $\eta$:
        \begin{equation}
            C_{s\to k} = \log \left(  1 +  \frac{P_s |h_{s\to k}|^2}{\sigma^2} \right) > \eta
        \label{eq:eC1}
        \end{equation}
        where $\sigma^2$ is the noise power.
    \item At least the energy storage of one of the non-selected relays can be charged enough to transmit one packet or more:
    \begin{equation}
        \Delta E_{i} \geq E_r, \qquad~ \exists i \neq k
    \label{eq:eC2}
    \end{equation}
\end{itemize}

By applying the above selection rule at the edge state, if $link_{s \to k}$ is selected for data transmission, the data buffer for $R_k$ will become full, and the energy storage for at least one of the other relays will be charged, allowing for data transmission in the next time slot.

\subsection{Non edge-states}

At non-edge states, since the system will not move to a deadlock in the next time slot, the selection rules prioritize maintaining as many available links as possible. To achieve this, higher selection priorities are given to links that are less likely to lead to empty or full data buffers and empty energy storage. The selection rules at the non-edge state $s(i)$ are as follows:
\begin{itemize}
    \item Set the priorities for all available links at $\s(i)$ as shown in Section \ref{se:linkpri}.
    \item Consider the available links one by one, from the highest to the lowest priorities.
    \item If the considered link capacity is higher than the target data rate $\gamma$, the link is selected for the data transmission. Otherwise, the link with the next higher priority is considered.
    \item Continue this process until a link is selected, or the outage occurs if no link can be selected.
\end{itemize}
It is important to note that the distinction between edge and non-edge states does not impact the priorities of the links.

\subsection{Implementation}

The primary cost of implementing the scheme, aside from data transmission, lies in setting the link priorities. These priorities are determined based on the \textit{availability indices}, as outlined in Table II. Calculating the availability indices requires minimal computational effort but demands a certain amount of memory. Specifically, in a network with $K$ relays, there are $2K$ links in total. For each selected link, three types of availability indices are generated for every relay node, as shown in \eqref{eq:mMsi}. Consequently, the total number of indices that need to be stored is $2K \times 3K = 6K^2$.

The link selection decision can be implemented either centrally or in a distributed manner. In the centralized approach, a central node (such as the source or destination relay node) utilizes backhaul links to gather channel state information and selects a link for data transmission. In the distributed implementation, each relay node has access to the instantaneous channel gains of its associated source-to-relay and relay-to-destination links, as well as the average channel gains of other links. At each time slot, every relay node performs the following steps:
\begin{itemize}
\item Monitor the data buffer lengths and energy storage levels of other relays to determine the current system state.
\item Evaluate the expected state availability vector for each associated link if it were selected, and determine the priority order for every link.
\item Assign a countdown clock to each associated link, where the clock duration is inversely proportional to the link's priority order.
\item When a countdown clock reaches zero, the corresponding link is used for transmission if it can support the target rate. The relay may also broadcast the status of its data buffer and energy storage to assist other relays in better estimating the system state. Upon detecting this transmission, all other relays will defer their own transmissions until the next time slot.
\end{itemize}

As demonstrated, each node can independently determine whether to transmit based on the instantaneous channel gains of its associated source-to-relay or relay-to-destination link, along with the average channel gains of other relays. This approach enables the implementation of the selection rules without requiring the continuous collection of instantaneous channel gains from all relays at a central node, thereby simplifying the process and reducing overhead.


\section{Outage analysis}
\label{sec:out}
We denote $\bf A$ as the $(L^{(D)}_{\max} + 1)^K \times (L^{(E)}_{\max} + 1)^K$-by-$(L^{(D)}_{max} + 1)^K \times (L^{(E)}_{\max} + 1)^K$ state transition matrix, where the entry $A_{ji} = P(\s(j)|\s(i))$ which is the transition probability from state $\s(i)$ to $\s(j)$. Particularly $A_{ii}$ is the outage probability at state $\s(i)$.  Below we derive the transition probability $A_{ji}$ for $i=j$ and $i \neq j$, from which the outage probability of the overall system is obtained.

We denote $g_k=|h_k|^2$ as the channel gain for $link_k$, which is exponentially distributed with average gain $\lambda_k$. The cumulative-distribution-function (CDF) of $g_k$ is given by
\begin{equation}
F_k(x) = 1- e^{-\frac{x}{\lambda_k}}
\label{eq:fkx}
\end{equation}
where the index `$k$' refers to either the ${s\to k}$ or ${k \to d}$. The following probabilities will be used in the later analysis.
\begin{itemize}
    \item The probability that $link_k$ is in outage is given by
    \begin{equation}
    \begin{aligned}
            P_k^{(o)} &= P(C_k < \gamma) \\
             &= P\left( \log\left( 1+ \frac{P_t g_k}{\sigma^2}  \right)  < \gamma \right)
    \end{aligned}
\label{eq:P1-0}
\end{equation}
where $\gamma$ is the target rate, and $P_t$ is either $P_s$ or $P_r$, for $link_{s\to k}$ and $link_{k\to d}$, respectively. Let $\xi =  (2^\eta -1)\sigma^2/P_t$, \eqref{eq:P1-0} becomes
 \begin{equation}
            P_k^{(o)} = P(g_k < \xi) = F_k(\xi)
        \label{eq:P1}
        \end{equation}

Correspondingly the probability that  $link_k$ is not in outage is given by
    \begin{equation}
    P_k^{(\bar{o})} = 1 -  F_k(\xi)
    \end{equation}

    \item The probability that the charging via $link_{s \to k}$ can increase the energy storage of $R_k$ by $m$ is given by
        \begin{equation}
            \begin{aligned}
            & P_{s \to k}^{(c)}(m) = \\
            & \left \{
             \begin{array}{l}
                 P(m E_t < \Delta E_{k} < (m+1) E_t), {\rm if~} 0 \leq m < L^{(E)}_{\max} - L^{(E)}_{k}\\
                 P(\Delta E_{k} > m E_t), ~~{\rm if~} m = L^{(E)}_{\max} - L^{(E)}_{k}
            \end{array}
            \right.
            \end{aligned}
        \label{eq:P2m0}
       \end{equation}
       where $\Delta E_{k}$ is given by \eqref{eq:DeltaE} which is the energy harvested at $R_k$. Applying \eqref{eq:DeltaE} and \eqref{eq:fkx} in \eqref{eq:P2m0}, we have
       \begin{equation}
          P_{s \to k}^{(c)}(m) = F_{s \to k}(x) - F_{s \to k} \left( m/\rho \right)
        \label{eq:P2m}
       \end{equation}
       where $x = (m+1)/\rho$ if $0 \leq m < L^{(E)}_{\max} - L^{(E)}_{k}$,  and $x \to \infty$ if $m = L^{(E)}_{\max} - L^{(E)}_{k}$.

    \item The probability that the charging via $link_{s \to k}$ can increase the energy storage of $R_k$ by $m$ and $link_{s \to k}$ is in outage is given by
     \begin{equation}
            \begin{aligned}
            & P_{s \to k}^{(c, o)}(m) = \\
            & \left \{
             \begin{array}{l}
                 P(g_{s \to k} < \xi, ~m E_t < \Delta E_{k} < (m+1) E_t), \\
                      \qquad \qquad \qquad \qquad \qquad  {\rm if~} 0 \leq m < L^{(E)}_{\max} - L^{(E)}_{k}\\
                 P(g_{s \to k}<\xi, ~\Delta E_{k} > m E_t, {\rm ~~if~} m = L^{(E)}_{\max} - L^{(E)}_{k}
            \end{array}
            \right.
            \end{aligned}
        \label{eq:P3m0}
       \end{equation}
    Similarly applying \eqref{eq:DeltaE} and \eqref{eq:fkx} in \eqref{eq:P3m0}, we have
    \begin{equation}
         P_{s \to k}^{(c, o)}(m) =
             \left \{
             \begin{array}{ll}
                 0, & {\rm if~} \xi < m \rho\\
                 F_{s \to k}(y) - F_{s \to k} \left( m/\rho \right), & {\rm if~} \xi > m \rho
            \end{array}
            \right.
              \label{eq:P3m}
    \end{equation}
    where $y = \min \{ (m+1)/\rho, ~\xi \}$ if $0 \leq m < L^{(E)}_{\max} - L^{(E)}_{k}$,  and $y = \xi$ if $m = L^{(E)}_{\max} - L^{(E)}_{k}$.

\end{itemize}


\subsection{Outage probability at state $s_i$: $P_{out}^{(\s(i))} = A_{ii}$}
\subsubsection{Edge-state}
Suppose $\s(i)$ is an edge-state and only the data buffer for relay $R_q$ is nearly full and all other data buffers are full. As is shown in \eqref{eq:eC1} and \eqref{eq:eC2}, outages occurs when $link_{s\to q}$ cannot support the target rate, or no other $link_{s\to k}$ ($k \neq q$) can charge the energy storage enough for future transmission.

The outage probability at edge state $\s(i)$ is then given by
\begin{equation}
\begin{aligned}
P_{out}^{(\s(i))} & = 1 - P_{s\to q}^{(\bar{o})} \cdot P({\rm at ~least ~one ~relay ~can ~be ~charged})   \\
&= 1 -  P_{s\to q}^{(\bar{o})}  \cdot \left( 1 - \prod_{k \neq q} P_{s \to k}^{(c)}(0) \right), \\
&= 1 -  (1-F_{s\to q}(\xi))   \cdot \left( 1 - F_{s\to k}(1/\rho)  \right),
\end{aligned}
\label{eq:Pouts}
\end{equation}
where, as is shown in \eqref{eq:P2m}, $ P_{s \to k}^{(c)}(0))$ is the probability that the energy harvested at relay $R_k$ cannot increase the energy storage from zero to one.

\subsubsection{non edge-state}

Suppose $\s(i)$ is a non-edge-state, and there are $N_i$ available links. The outage occurs if all of $N_i$ links are in outage. The outage probability is given by
\begin{equation}
\begin{aligned}
 P_{out}^{(\s(i))} &=  \prod_{k \in \mathbb{S}_{N_i}}  P_k^{(o)}
 = \prod_{k \in \mathbb{S}_{N_i}}  F_k(\xi)
 \end{aligned}
\label{eq:piiou}
\end{equation}
where $\mathbb{S}_{N_i}$ is the set of $N_i$ available links.


\subsection{Transition probability $A_{ji} = P(\s(j)|\s(i), i \neq j$)}

As is shown in \eqref{eq:std}, a state consists of two parts: data state $\s_d$ and energy states $\s_e$, respectively. The transition probability can be expressed as
\begin{equation}
     P(\s(j)|\s(i)) =  P(\s^{(D)}(j)|\s^{(D)}(i)) \cdot P(\s^{(E)}(j)|\s^{(E)}(i))
\end{equation}

Below we derive data state transition $P(\s^{(D)}(j)|s^{(D)}(i))$ and energy state transition $P(\s^{(E)}(j)|s^{(E)}(i))$, respectively. Without losing generality, we assume $link_q$ is selected for $\s(i)$ being transited to $\s(j)$.

\subsubsection{Data state transition $P(\s^{(D)}(j)|\s^{(D)}(i))$}
When $link_q$ is selected for data transmission, the data buffer of $R_k$ is either increased or reduced by one, if $link_q$ is $link_{s\to q}$ or $link_{q\to d}$, respectively. The data buffer for all other relays remain unchanged. Therefore, when the state transits from $\s(i)$ to $\s(j)$, the data state must be changed (i.e. $\s^{(D)}(j) \neq \s^{(D)}(i)$).

Because different selection rules are applied to edge-state and non edge-states, the corresponding transition probabilities also have two forms as follows:

{\em Case 1: $\s(i)$ is an edge-state}. In this case, the only available link is $link_{s \to q}$ which is selected if there is no outage at this state. The transition probability is given by
\begin{equation}
    P(\s^{(D)}(j)|\s^{(D)}(i)) = 1 - P_{out}^{(\s(i))},
\end{equation}
where $P_{out}^{(\s(i))}$ is obtained in \eqref{eq:Pouts}.

{\em Case 2: $\s(i)$ is a non edge-state}. In this case, from the selection rules, $link_q$ can only be selected if it can support the target data rate and all other links with higher priority are in outage. Thus we have
\begin{equation}
    P(\s^{(D)}(j)|\s^{(D)}(i)) = P_q^{(\bar{o})} \prod_{k: {\mathcal R}(k) > {\mathcal R}(q)} P_k^{(o)},
\end{equation}
where ${\mathcal R}(k) > {\mathcal R}(q)$ states that $link_k$ has higher priority than $link_q$.

\subsubsection{Energy state transition $P(\s^{(E)}(j)|\s^{(E)}(i))$}

When the state transits from $\s(i)$ to $\s(j)$, unlike the data state which must be changed, the energy state may either transit to another one or remain unchanged, This depends on how much energy is harvested at the corresponding relays.

Note that the energy state vector is given by
\begin{equation}
    \s^{(E)}(x) = [L^{(E)}_{1}(x), \cdots, L^{(E)}_{K}(x)],
\label{eq:SExx}
\end{equation}
where $L^{(E)}_{k}(x)$ is the length of the $k$-th energy storage at state  $\s^{(E)}(x)$. When energy state transits from $\s^{(E)}(i)$ to $\s^{(E)}(j)$, the change of the energy storage at the relay $R_k$ is denoted as
\begin{equation}
    m_k = L^{(E)}_{k}(j) - L^{(E)}_{k}(i), \qquad k= 1, \cdots, K
\label{eq:mfL}
\end{equation}

Depending on whether $link_q$ is a $S \to R$ or $R \to D$ link, we have the following two cases for the energy state transition probability:

{\em Case 1:} The selected link is $link_{q \to  d}$, a relay-to-destination link. In this case, the energy storage of relay $R_q$ is decreased by one, and all other energy storage remain unchanged. In other words, we have $m_q = -1$, and $m_k=0$ for all $k \neq q$. The energy state transition probability is given by
\begin{equation}
    P(\s^{(E)}(j)|\s^{(E)}(i))=1
\end{equation}

{\em Case 2:} The selected link is  $link_{s \to q}$, a source-to-relay link. In this case, $link_{s \to q}$ is used for data transmission, and all other $link_{s \to k}$ links are charging the corresponding relay $R_k$. Because all links are independently fading, from \eqref{eq:SExx} we have
\begin{equation}
    P(\s^{(E)}(j)|\s^{(E)}(i)) = \prod_{k: k \neq q}  P(L^{(E)}_k(j)|L^{(E)}_k(i)).
\label{eq:Pseji}
\end{equation}

In order for the energy state to transit from $\s^{(E)}(i))$ to $\s^{(E)}(j)$, the energy harvesting at relay $R_k$ needs to increase the energy storage length by $m_k$, where $m_k$ is defined in \eqref{eq:mfL}.

If $link_{s \to k}$ has lower priority than $link_{s \to q}$, i.e. ${\mathcal R}(k) < {\mathcal R}(q)$, from \eqref{eq:P2m} the probability for the energy storage at $R_k$ increased by $m_k$ is given by
\begin{equation}
     P(L^{(E)}_k(j)|L^{(E)}_k(i)) = P^{(c)}_{s\to k}(m_k), \quad {\rm if~} {\mathcal R}(k) < {\mathcal R}(q)
\label{eq:ple1}
\end{equation}

On the other hand, if $link_{s \to k}$ has higher priority than $link_{s \to q}$, its line capacity must be smaller than the target rate as otherwise it will be selected for data transition. Therefore from \eqref{eq:P3m} we have
\begin{equation}
     P(L^{(E)}_k(j)|L^{(E)}_k(i)) = P^{(c, o)}_{s\to k}(m_k), \quad {\rm if~} {\mathcal R}(k) > {\mathcal R}(q)
\label{eq:ple2}
\end{equation}

Substituting \eqref{eq:ple1} and \eqref{eq:ple2} into \eqref{eq:Pseji}, gives
\begin{equation}
\begin{aligned}
P(\s^{(E)} & (j)|  \s^{(E)}(i))  =\\
& \prod_{k: {\mathcal R}(k) < {\mathcal R}(q)}   P^{(c)}_{s \to k}(m_k) \cdot  \prod_{k: {\mathcal R}(k) > {\mathcal R}(q)}   P^{(c,o)}_{s \to k}(m_k)
\end{aligned}
\label{eq:Psfin}
\end{equation}

\subsection{Overall outage probability}

The outage probability of the overall system is given by
\begin{equation}\label{eq:pout}
P_{out} = \sum_{i=1}^{L_{s}} \pi_i\cdot p_{out}^{(\s(i))},
\end{equation}
where $L_s=[(L^{D}_{\max}+1)\times (L^{E}_{\max}+1) ]^K$ which is the total number of states, $\pi_i$ is the stationary probability for state $\s_i$.

From above, we can obtain every entry of the transition matrix $\mathbf A$. Because the transition matrix $\bf A$ is column stochastic, irreducible and aperiodic, the stationary state probability vector is obtained as (see \cite{J.R98,A.B94})
\begin{equation}
\bfg \pi=(\textbf{A}-\textbf{I}+\textbf{B})^{-1}\textbf{b}
\label{eq:bpi}
\end{equation}
where $\bfg \pi = [\pi_1, \cdots, \pi_{(L+1)^N}]^\tra$, $\textbf{b} = (1, 1,..., 1)^T$, $\textbf{I}$ and $\textbf{B}$ are the identity and all one matrices with appropriate dimensions respectively.

Substituting \eqref{eq:bpi} into \eqref{eq:pout} gives the overall outage probability as
\begin{equation}\label{eq:Poutge}
\begin{aligned}
P_{out} = \textrm{diag}(\textbf{A})^\tra \cdot (\textbf{A}-\textbf{I}+\textbf{B})^{-1}\textbf{b},
\end{aligned}
\end{equation}
where $\textrm{diag}(\textbf{A})$ is a vector consisting of all diagonal elements of $\bf A$.


\subsection{Diversity order}
\label{sec:div}

The diversity order reveals the average number of available links for selection for very high signal-to-noise ratio (SNR), which shows the best potential performance of the network. We define the diversity order as
\begin{equation}
\begin{aligned}
d &= - \lim_{P_s \to \infty} \frac{\log P_{out}}{\log P_s}  \\
    &= - \lim_{P_s \to \infty} \sum_{i=1}^{L_{s}} \pi_i \frac{\log p_{out}^{(\s(i))}}{\log P_s}
\end{aligned}
\label{eq:dDef}
\end{equation}
It is clear that the SNR goes to infinity when $P_s \to \infty$. From \eqref{eq:P1} and \eqref{eq:fkx} we have
\begin{equation}
    \lim_{P_s \to \infty} P_k^{(o)} = 0
\end{equation}
Thus no link is in outage when $P_s \to \infty$.

To achieve a full diversity order of $2K$, the minimum size of the data buffer should be three, allowing the buffer length to stay at one or two to avoid being empty or full. Similarly, the minimum energy storage size should be two, so the energy storage length can be one or two to avoid being empty.

Without loss of generality, we consider a system with the minimum buffer and energy sizes, assuming that the system starts with empty buffers and energy storage. Because no link is in outage when $P_s \to \infty$, initially a source-to-relay link will be selected. This will increase the corresponding buffer length by one and charge all other energy storage units. At this stage, relay-to-destination links have lower priorities than source-to-relay links. According to the proposed selection rules, if there are enough relays, source-to-relay links will continue to be selected until all data buffer lengths are two, with most energy storage units charged to two and the rest to one.

Subsequently, relay-to-destination links with energy storage levels of two will be prioritized for selection. This action will reduce the lengths of the corresponding data buffers and energy storage units to one. Following this, the source-to-relay links will be selected once again. This iterative process will continue until all data has been successfully transmitted. As a result, in steady-state conditions, there will be no empty or full data buffers, and no energy storage units will be depleted. This ensures that the number of available links remains consistently at $2K$. To maintain this balance, a sufficiently large number of relays is required, providing ample opportunities for the relays to charge. The specific number of relays needed to achieve full diversity depends on several factors, including the harvesting coefficient $\rho$, as well as the sizes of the data buffers and energy storage units.

Therefore, from \eqref{eq:piiou}, the outage probability at any state, $s(i)$, is given by

\begin{equation}
 P_{out}^{(\s(i))} =  \prod_{k=1}^{2K} F_k(\xi)
 =\left( 1 - e^{-\frac{\xi}{\lambda}} \right) ^{2K}
\label{eq:pihs}
\end{equation}
 Substituting \eqref{eq:pihs} into \eqref{eq:dDef}, and further noting that $\lim_{P_s \to \infty } e^{-\xi/\lambda} = 1- \xi/\lambda$, we have
\begin{equation}
\begin{aligned}
    d &= - \lim_{P_s \to \infty} \frac{\log p_{out}^{(\s(i))}}{\log P_s} \\
    &= - \lim_{P_s \to \infty} \frac{2K(\log \xi - \lambda)}{\log P_s}
    &= 2K
\end{aligned}
\label{eq:dreS}
\end{equation}

The relay power coefficient $\alpha$ is assumed constant above. However, when the source transmission power $P_s$ increases, $\alpha$ can be smaller, facilitating easier charging. As $P_s \to \infty$, $\alpha$ can become arbitrarily small. Consequently, only a small number of relay nodes are needed to achieve full diversity.


\section{Numerical Simulations}
\label{sec:num}

Numerical results are shown in this section to verify the proposed scheme. In all simulations below, the average channel gains for all channels are normalized to one. The SNR(dB) in the figures is defined as $10\log(P_s/\sigma^2)$, where $P_s$ is the transmit power at the source and $\sigma^2$ is the noise power.

\begin{figure}[t!]
  \centering
  \centerline{\includegraphics[scale=0.63]{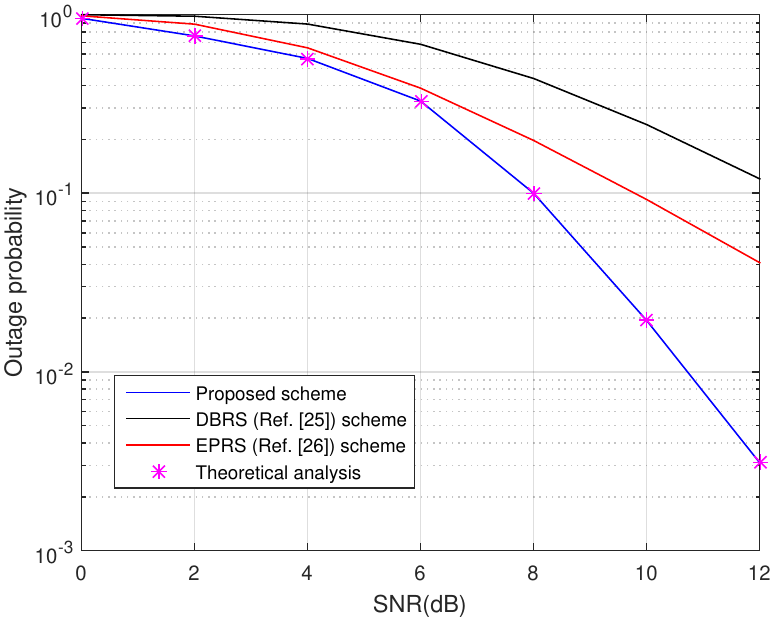}}
 \caption{\small Theoretical vs simulation results with $K=2$, $\rho=0.5$, $L_{\max}^{(D)}=3$, $L_{\max}^{(E)}=2$ and $\alpha=1$.} \label{fig:theo}
\end{figure}

Figure \ref{fig:theo} compares the theoretical and simulation analyses for a system with two relays ($K=2$), a data buffer size of  $L_{\max}^{(D)}=3$, an energy storage size of $L_{\max}^{(E)}=2$, a charging coefficient of $\rho=0.5$, and a relay energy coefficient of  $\alpha=1$, resulting in $P_t = \alpha P_r = P_r$. The theoretical results align closely with the simulation results. For comparison, the performance of the DBRS scheme proposed in \cite{LiuK2021} and the EPRS scheme \cite{HuangS2022} are also included. It is important to note that both the DBRS scheme in \cite{LiuK2021} and the proposed scheme in this paper assume no inter-relay charging, whereas the EPRS scheme in \cite{HuangS2022} assumes inter-relay charging. As clearly demonstrated in Fig. \ref{fig:theo}, the proposed scheme significantly outperforms the DBRS scheme. For example, an outage probability of 0.1 is achieved at an SNR of 8 dB with the proposed scheme, compared to over 12 dB with the DBRS scheme. Additionally, it is noteworthy that, even with inter-relay charging, the EPRS scheme \cite{HuangS2022} still underperforms compared to our proposed scheme. This outcome is expected because both the DBRS and EPRS schemes are constrained to follow a fixed alternating order of source-to-relay and relay-to-destination This limitation restricts the full potential of buffer-aided relay networks. In contrast, the proposed scheme relaxes this constraint by allowing the selection of a link from all available links, thereby enhancing flexibility and performance.

\begin{figure}[t!]
  \centering
  \centerline{\includegraphics[scale=0.7]{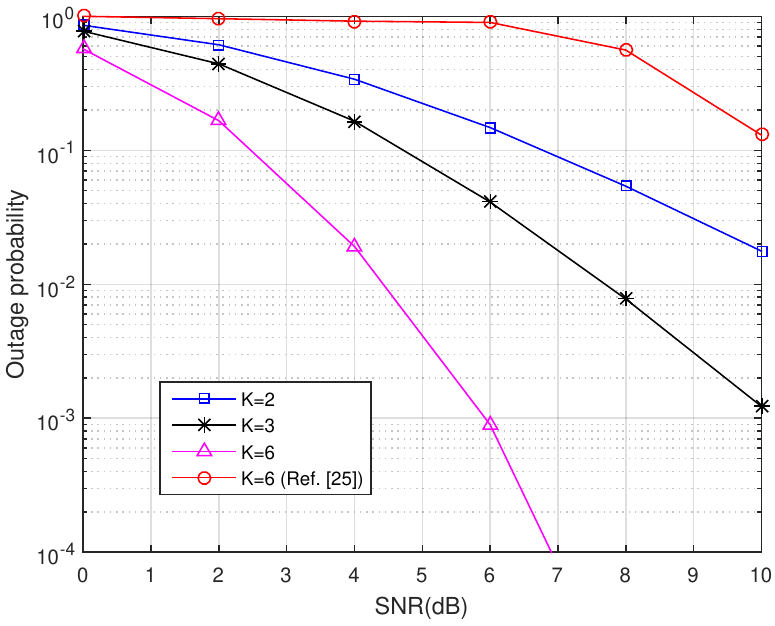}}
 \caption{\small Outage probability vs SNR for different number of relays, where $\rho=0.5$, $L_{\max}^{(D)}=3$, $L_{\max}^{(E)}=2$, and $\alpha=1$.}
\label{fig:relay}
\end{figure}

Figure \ref{fig:relay} illustrates the outage probability of the proposed scheme for varying numbers of relays, where all parameters remain consistent with those in Figure \ref{fig:theo}, except for the number of relays $K$. For comparison, the results of the DBRS scheme \cite{LiuK2021} with $K=6$ are also included. It is evident that increasing the number of relays improves the outage performance. Specifically, the full diversity order of $2K=12$ can be observed in the proposed scheme with $K=6$, as determined by measuring the gradient of the curve at high SNR values. To further clarify the diversity order, Table \ref{tab:div} presents the approximate diversity order of the proposed scheme, measured at SNR=7dB, and that of the DBRS scheme, measured at SNR=17dB. The results align well with the analysis in Section \ref{sec:div}, demonstrating that when the number of relays is sufficiently large, the proposed scheme achieves the full diversity order of $2K$, whereas the DBRS scheme only achieves half of the full diversity order, $K$. This difference arises because the proposed scheme selects a link from all available links, with a maximum of $2K$ options, while the DBRS scheme selects a link from either the maximum $K$ source-to-relay or relay-to-destination links, limiting its maximum diversity order to half of $2K$.

\begin{table}[htbp]
\caption{Diversity orders comparison with $K=6$}
\label{tab:div}
\centering
\resizebox{\columnwidth}{!}{
\begin{tabular}{|c|c|c|c|}

     \hline
             & $P_{out}(SNR)$ (dB)& $P_{out}(SNR)$ (dB)& Diversity order \\
    \hline
    Proposed & $P_{out}(6~{\rm dB} )=30$ & $P_{out}(7~{\rm dB} )= 41.55$ & $\frac{41.55-30}{7-6} \simeq 12$  \\
    \hline
    Ref. \cite{LiuK2021} & $P_{out}(16~{\rm dB} )=29.59$ & $P_{out}(17~{\rm dB} )=35.53$ & $\frac{35.53-29.59}{17-16} \simeq  6$  \\
    \hline
\end{tabular}%
}
\end{table}

\begin{figure}[t!]
  \centering
  \centerline{\includegraphics[scale=0.7]{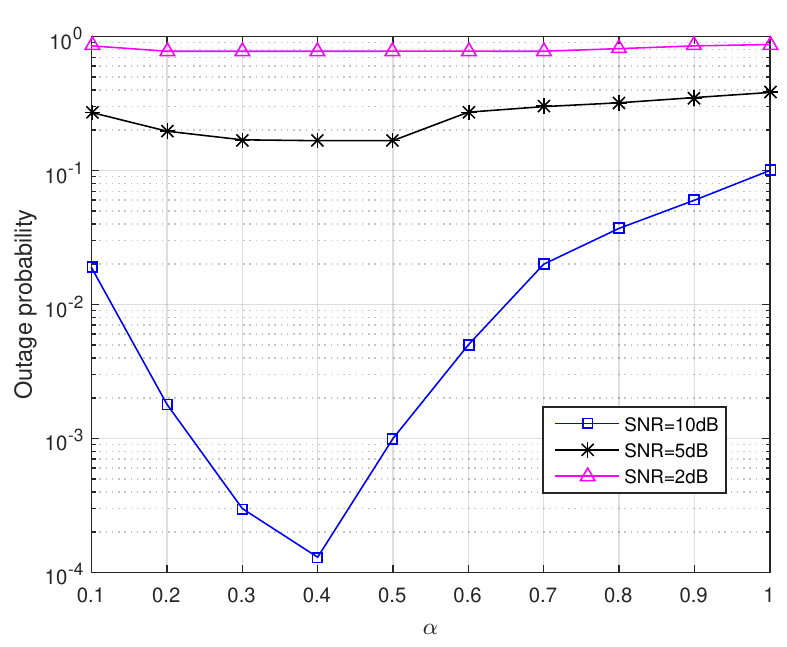}}
 \caption{\small The influence of relay power coefficient $\alpha$, where $K=6$,  $L_{\max}^{(D)}=3$, $L_{\max}^{(E)}=10$ and $\rho=0.5$} \label{fig:alpha}
\end{figure}

Fig. \ref{fig:alpha} shows how the outage performance is influenced by the relay power coefficient $\alpha$. It is known that a higher $\alpha$ improves the relay-to-destination transmission due to the higher transmit power applied at the relays. However, a higher $\alpha$ also makes it more difficult for the relay to harvest wireless energy. Generally, the higher the transmission power $P_s$, the smaller the $\alpha$ required. This is clearly demonstrated in Fig. \ref{fig:alpha}. Notably, the outage performance is more sensitive to $\alpha$ at higher SNRs. This is because, at high SNR, energy harvesting has a greater impact on the system than the transmit powers at the relays.

\begin{figure}[t!]
  \centering
  \centerline{\includegraphics[scale=0.7]{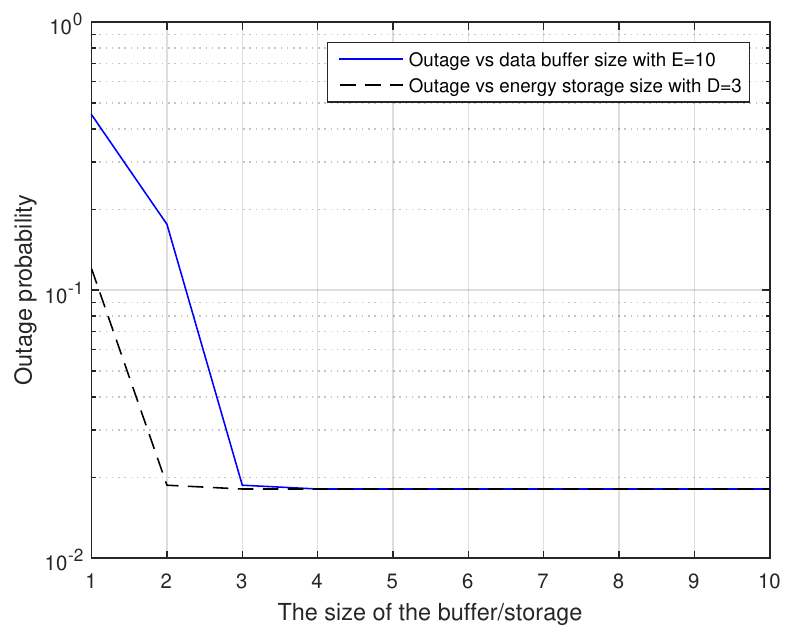}}
 \caption{\small The influence of the data buffer size and the energy storage size, where $K=3$, $\rho=0.5$, average channel gain 0.5, $\alpha=0.4$ and SNR=10.} \label{fig:buffer}
\end{figure}

Fig. \ref{fig:buffer} illustrates the impact of data buffer and energy storage sizes on the outage probability. The solid line curve represents the outage probability as a function of the data buffer size, with the energy storage size fixed at $E=10$. The dotted line curve, on the other hand, depicts the outage probability as a function of the energy storage size, while the data buffer size is held constant at $D=3$. It clearly demonstrates that increasing the size of either the data buffer or energy storage improves outage performance, but the improvement becomes less significant once the size is sufficiently large. Interestingly, increasing the data buffer size appears to have a greater impact on reducing the outage probability than increasing the energy buffer size. For example, with a data buffer size of  $L_{\max}^{(D)}=1$ and an energy storage size of $L_{\max}^{(E)}=10$, the outage probability is above 0.45. In contrast, with $L_{\max}^{(D)}=3$  and $L_{\max}^{(E)}=1$, the outage probability is 0.12. This is because energy harvesting depends on the channel strength which is not always strong. Thus it is usually not easy to fully charge an energy storage, making the large energy size less necessary.

\section{Conclusion} \label{sec:con}
This paper proposes a novel selection scheme for relay networks equipped with both data buffers and energy storage. The closed-form expression of the outage probability for the proposed scheme has been derived, and the diversity order has been analyzed. Numerical simulations have been conducted to verify the proposed scheme. Both the analysis and simulations show that the proposed scheme can achieve a full diversity order of $2K$. The proposed scheme can be implemented in a distributed manner, allowing each relay node to make its own transmission decisions. This is in contrast to many other relay selection schemes that require gathering instantaneous channel coefficients of all links at a central node. This paper focuses on the outage performance of the network. Future work may include incorporating other performance metrics, such as network delay, into the selection rule.

\balance
\bibliographystyle{ieeetr}
\bibliography{ref}


\end{document}